\documentclass[12pt]{JHEP3} 
\usepackage{amsfonts}   
\usepackage{amsmath}   
\usepackage{multicol}   
\usepackage{epsfig}   
\usepackage{graphicx}   
    
\font\myb=msbm10 at 12pt
\def\bb#1{\hbox{\myb#1}}

\def\P{\Pi\!\!\!\!/}\def\cC{{\cal C}}
\def\a{\alpha} \def\b{\beta} \def\g{\gamma} \def\l{\lambda} \def\d{\delta} \def\e{\epsilon} \def\t{\theta}  
\def\s{\sigma}   \def\G{\Gamma}  \def\n{\nu} 
  \def\p{\partial}

\newcommand{\be}{\begin{equation}} 
\newcommand{\ee}{\end{equation}}
\newcommand{\bea}{\begin{eqnarray}} 
\newcommand{\eea}{\end{eqnarray}}
\newcommand{\nn}{\nonumber}

\title{\Large Super D-branes from BRST Symmetry}   
\author{Lilia Anguelova 
\\ C.N. Yang Institute for Theoretical Physics,  SUNY at Stony Brook\\ 
Stony Brook, NY 11794-3840, USA, \\
anguelov@insti.physics.sunysb.edu}   
\author{Pietro Antonio Grassi\\ C.N. Yang Institute for Theoretical Physics,  
SUNY at Stony Brook\\ 
Stony Brook, NY 11794-3840, USA, \\
pgrassi@insti.physics.sunysb.edu}   
   
\abstract{   
Recently a new formalism has been developed for the covariant quantization of superstrings. We study properties of D$p$-branes and $p$-branes in this new framework, focusing on two different topics: effective actions and boundary states for D$p$-branes. 
We present a derivation of the Wess-Zumino terms for super (D)$p$-branes using BRST symmetry. To achieve this we derive the BRST symmetry for superbranes, starting from the approach with/without pure spinors, and completely characterize the WZ terms as elements of the BRST cohomology. We also develope the boundary state description of D$p$-branes by analyzing the boundary conditions for open strings in the completely covariant (i.e., without pure spinors) BRST formulation.
}

\preprint{     
YITP-03-35
}      
     
\begin{document}



\section{Introduction and summary} 

One of the most important problems in string theory is the covariant quantum formulation of the 
target-space supersymmetric Green-Schwarz superstring \cite{superstring}. It is highly nontrivial because of the entanglement between first and second class fermionic constraints. Equivalently, the difficulties are due to the $\kappa$-symmetry \cite{siegel}, that the GS superstring possesses, which is infinitely reducible and hence leads to an infinite tower of ghost fields. Nevertheless, $\kappa$-symmetry is fundamental 
to achieve supersymmetry on the worldvolume of (D)$p$-branes and in fact, it is even sufficient to determine the structure of their (effective) actions: 
the kinetic term $S_K$, which is the Dirac-(Born-Infeld) action, has to be supplemented 
by a Wess-Zumino term $S_{WZ}$ for $\kappa$-invariance \cite{pbrane,curvKappaAc,APS}. 

On the other hand, recently a new approach to the covariant quantization of superstrings 
has been developed. It is based on the pure spinor formalism \cite{berko} 
and it has been extended in \cite{GPPN, grassi} to encode all the constraints into BRST symmetry. 
The key idea is to start from a quantizable action and 
to construct the correct BRST differential whose cohomology describes the 
spectrum of the superstring. In this way, $\kappa$-symmetry 
is replaced by a BRST charge. This suggests that BRST symmetry should 
also characterize the (effective) actions for (D)$p$-branes. 
  
We will show that this is indeed the case. Namely, we will find that the kinetic term in the action of a (D)$p$-brane determines BRST transformation rules which allow us to extract the Wess-Zumino (WZ) term as a uniquely selected nontrivial cocycle of a suitable cohomology group. The BRST differential $s$ and the worldvolume exterior derivative $d$ form a double complex and one can write descent equations for any cohomology group $H^{r}_{q}(s|d)$, where the lower index $q$ is the worldvolume form number and the upper index $r$ - the ghost number. The WZ terms are non-trivial cocycles of these cohomologies. In the Berkovits' formalism the chain of descent equations stops at the third level, namely at $H^{2}_{p}$ (where $p$ is the number of worldvolume spatial dimensions), and then it is difficult to prove the uniqueness of $S_{WZ}$. However, by a suitable extension of the pure spinor formalism, that relaxes all constraints, one can construct a representative for each cohomology group of the series $H^{k}_{p+2-k}(s|d)$ with $0 \leq k \leq p+2$. In this case the WZ term is uniquely determined by the highest ghost number cohomology group.

Once we have found $S_{WZ}$, we can study how the BRST charge and transformation rules are modified by it. We will see that in order to obtain a nilpotent BRST charge one has to impose new constraints on the commuting spinors. These constraints are related to new ghost fields. It turns out that the new transformation rules can be obtained by generalizing the construction of \cite{grr,branealgebra, branealgebra1} (based on the analysis of the Chevalley-Eilenberg (CE) cohomology \cite{CE} of a suitably extended superspace) to a BRST differential $s + d$, first introduced in \cite{bal}. 
More precisely, the authors of \cite{branealgebra, branealgebra1} showed that for super (D)$p$-branes the usual super-Poincar\'e algebra is extended with new generators to a so called brane superalgebra. This yields an enlarged group manifold, for each new coordinate of which according to our approach one has to introduce a new ghost field. The full set of ghosts forms irreducible supermultiplets and the complete BRST transformation rules are obtained via a simple prescription from the CE cohomology.
  
As is well-known, D$p$-branes are not fundamental objects in string theory, but solitons whose small 
excitations are described by collective coordinates. Hence the Dirac-Born-Infeld action is merely an effective action 
for these coordinates implementing the dynamics of superstrings 
ending on the branes. Nevertheless, this effective action can be viewed as describing a field theory living on the brane worldvolume, whose gauge symmetries are worldvolume diffeomorphisms, $\kappa$-symmetry and a $U(1)$ gauge 
symmetry. In the present framework, we have replaced all these symmetries by BRST 
symmetry\footnote{Replacing all the gauge symmetries of a system with a global BRST symmetry (fermionizing all 
the gauge parameters) seems to suggest that the resulting action could be derived from a gauge invariant 
action by a gauge-fixing procedure. The natural candidate for such gauge invariant action is the 
$\kappa$-invariant $Dp$-brane action. However, it is not clear how to derive the BRST invariant action 
from the latter. The superembedding formalism 
(see for example \cite{tonin} and \cite{sorokin}) might be a way 
to define a suitable gauge-fixing procedure.  
} 
and we have derived the same properties that follow from 
$\kappa$-symmetry. Although in the case of strings this approach has been very successful, the quantization of $D$-branes is rather involved due to the fact that  the Hamiltonian is not quadratic (see for example 
\cite{kallosh} and references therein). 
Some aspects of this quantization will be published separately. 
  
In a somewhat different direction, we have studied in the second part of this paper D-branes from the perspective of open superstrings in the BRST formulation of \cite{GPPN}. As known since the work of \cite{CNLY1, CNLY2, PC}, boundary conditions for open strings can be solved in operator form in terms of boundary states. The latter were introduced in order to facilitate the calculation of one loop diagrams in open string theory in the NSR formulation. The idea is to use worldsheet duality according to which an open string going around a loop can be viewed as a closed string created from the vacuum at some moment of worldsheet time $\tau_i$, propagating till some other moment $\tau_f$ and then disappearing into the vacuum again. Hence one can calculate the open string one loop amplitude by sandwiching a closed string propagator between two boundary states $|{\rm B}>$ and $<{\rm B}|$ which describe the creation from and the annihilation into the vacuum of a closed string respectively. When the open strings under consideration are constrained to end on fixed subspaces of the target space, the corresponding boundary states provide description of D-branes as coherent states of closed strings \cite{Li, CK}. In this case the above prescription for sandwiching of a closed string propagator between two boundary states gives the static interaction between the corresponding D-branes. 

Motivated by this, we have analyzed the boundary conditions for open strings in the completely covariant formulation of \cite{GPPN}. These conditions are determined by supersymmetry, BRST invariance and ghost number conservation. Having found them, we are able to write down the D-brane boundary states they lead to. This is the first step towards open string one loop computations in this formalism although still a lot of work remains to be done.

The present paper should be regarded as initiating the analysis of D$p$-branes in the new framework of covariant 
quantum superstrings.\footnote{We should mention, though, the derivation of the Born-Infeld action from open 
superstrings in the pure spinor formulation \cite{BP}.} There are still several open issues as well as many applications to be studied. 
One of the most important unsolved questions is the uncovering of the entaglement between BRST symmetry in the approach with/without pure spinors and the gauge symmetries of the 
classical actions. In particular, the role of diffeomorphisms in the pure spinor fromalism is not yet understood and therefore computations at higher orders in the string coupling are not yet feasible. On the other hand, there is a very promising recent development in the formulation without pure spinors \cite{superWZW}, namely an $N=2$ superconformal algebra was discovered which contains the quartet of topological gravity. Such structure has been found in all known so far covariant formulations of superstrings in different dimensions \cite{TopStr} and it greatly facilitates the calculation of amplitudes.

The paper is divided in two parts: In the first one, Section 2, we discuss the (effective) actions for (D)$p$-branes. In Subsection 2.1 we extract a preliminary form of BRST symmetry from the kinetic term of the action. In Subsection 2.2 we use this pre-BRST symmetry to derive the WZ term in a formulation without constraints on the commuting spinor ghosts. In Subsection 2.3 we explore the extended BRST symmetry due to the adition of the WZ term. In Subsection 2.4, using a generalized form of the CE cohomology, we relate our BRST symmetry considerations with the brane superalgebras of \cite{branealgebra, branealgebra1}. 
In the second part, Section 3, 
we study the boundary conditions for completely covariant open superstrings and write down the corresponding D-brane boundary states. In Subsection 3.1 some preliminaries are discussed. In Subsection 3.2 the boundary conditions at fixed worldsheet time $\tau$ are derived. For completeness in Subsection 3.3 we also find the boundary conditions at fixed $\sigma$. Finally, in Subsection 3.4 we construct the boundary states.
  
\section{Wess-Zumino terms}

\subsection{BRST symmetry from the kinetic terms} 

The action of an extended ($p+1$-dimensional) bosonic object (brane) has the standard form\footnote{For a fundamental object (\ref{kin}) is {\it the} action whereas for a solitonic one $S_K$ provides a low energy effective description.}
\be\label{kin}
S_{K} = - \int d^{p+1} \sigma \sqrt{ - \det {\cal M}}
\ee
where ${\cal M}$ is the metric $g_{\mu \nu} = \eta_{mn} \p_{\mu} x^{m }\p_{\nu}  x^{n}$ induced by the embedding of the brane in the target space.\footnote{We denote by $\mu, \nu, \dots$ the worldvolume vector indices, by $m,n, \dots$ the target space 
vector indices and by $\a, \b, \g, \dots$ the target space spinor indices.} In the case of super $p$-branes there are two basic superinvariants
\be\label{suin}
\Pi^{m} = d x^{m} + \bar{\t}^{\a} \Gamma^{m}_{\a\b} d\t^{\b}\,, ~~~~~~  \Pi^{\a} = d \t^{\a}
\ee
and ${\cal M}$ is the supersymmetric generalization of the induced metric, namely $G_{\mu \nu} = \eta_{mn} \Pi_{\mu}^{m } \Pi_{\nu}^{n}$. The fermions $\t^{\a}$ are in the smallest representation of the double cover of the target space Lorentz group. Clearly now $S_{K}$ generalizes to higher dimensions the Nambu-Goto action for superstrings. 

For D$p$-branes the kinetic term has to be modified to take into account the 
world\-volume gauge field $A_{\mu}$. The latter appears only in the supersymmetric combination \cite{Town}:
\be\label{gauge}
{\cal F} = F - b\,, ~~~~~~ b = - \bar{\t} \G^{\sharp} \G_{m} d\t \Big( d x^{m} + {1\over 2} \bar{\t} \G^{m} d \t \Big) \,,
\ee
where $F$ is the field strength of $A_{\mu}$ and $\G^{\sharp}$ denotes either $\G^{11}$ for type IIA or $\tau_{3}$ for type IIB. The matrix ${\cal M}$ in (\ref{kin}) is now given by the linear combination $G_{\mu\nu} + {\cal F}_{\mu\n}$. The spinors $\t^{\a}$ are again in a suitable representation for target space supersymmetry. This means that in the case of type IIA theories (containing D$p$-branes with $p$ = even) $\t^{\a}$ 
is a 32 component Majorana spinor, which can be decomposed in two opposite chirality ones ${1\over 2} (1 \pm  \G^{11}) \t$, whereas for type IIB branes there are two Majorana-Weyl fermions $\t^{i \a}$ with the same chirality which belong to an $SU(2)$ doublet.\footnote{In the following the index $i = 1,2$ will be suppressed for convenience.}

From the kinetic term (\ref{kin}) it is possible to 
derive a preliminary form of the BRST symmetry (a pre-BRST symmetry) that will be sufficient for 
the purposes of this section. Differentiating the lagrangian $S_{K} = \int {\cal L}_{K}$ with respect to $\p_{0}\t^{\a}$, one finds \cite{hatsuda}: 
\be\label{ferm}
d_{\a} \equiv p_{\a} + P_{m} (\Gamma^{m} \t)_{\a} +  
E^{i} \Big[ (\Gamma^{m } \Gamma^{\sharp} \t)_{\a} ( \Pi_{i m} + {1\over 2} \bar{\t} \Gamma_{m} \p_{i} \t) 
- {1\over 2} (\Gamma^{m}\t)_{\a} (\bar{\t} \Gamma_{m} \Gamma^{\sharp} \p_{i} \t) \Big] ,
\ee
where we have introduced the conjugate momenta $P_{m}$, for the bosonic coordinates $x^{m}$, and 
$E^{i}$, for the spatial components of the gauge field $A_{\mu}$ (recall that, as is well-known, the 
temporal component $A_0$ has vanishing conjugate momentum). Finally, $p_{\a}$ is the momentum 
conjugate to $\t^{\a}$. Note that we cannot impose the 
fermionic constraint $d_{\a} =0$ since in the calculation of $d_{\a}$ we have not used the complete action. More precisely, we have not taken into account the Wess-Zumino term $S_{WS}$ needed for invariance under $\kappa$-symmetry. This is a key point for the following considerations. Namely, we will show that one can {\it find} the WZ term starting from the kinetic one and using only BRST symmetry.

We want to draw the reader's attention to the fact that from the point of view of the above derivation, based on \cite{hatsuda}, the right-hand side of (\ref{ferm}) is just a part from the full fermionic constraint. However, from the perspective of the BRST formulation the expression in that constraint (i.e., $d_{\a}$) is viewed as a field. Hence in the BRST approach one has to add to the action the term $S_{gf} = \int d^{p+1} \sigma d_{\a} \dot\t^{\a}$, which breaks $\kappa$-symmetry, in order to obtain (\ref{ferm}) as an equation rather than a definition for $d_{\a}$. In \cite{tonin} it has been suggested that $S_{gf}$ can be viewed as a gauge fixing term 
for $\kappa$-symmetry and cosequently it has been proposed that the quantum action 
has the form $S_{K} + S_{WZ} + Q (\int d^{p+1} w_{\a} \dot\t^{\a})$, where $w_{\a}$ is the conjugate momentum for a new commuting spinor field $\l^{\a}$.

With the help of this spinor $\l^{\a}$ we construct a BRST charge via the ansatz
\be\label{BRST}
Q = \int d^{p}\sigma \bar\l^{\a} d_{\a}\,,
\ee
as proposed by Berkovits \cite{berko}. In order to study the 
nilpotency of this BRST charge, we need to compute the Poisson brackets of the 
fermionic constraints, in our case the pre-constraints $d_{\a}$ in (\ref{ferm}). Making use of the disscusion in \cite{hatsuda} 
we obtain
\be\label{nilpo}
\{Q, Q\} = 
\int d^{p}\sigma  \Big( \bar{\l}^{\a} \Xi_{\a\b} \l^{\b} +  (\p_{i} E^{i}) \bar{\l}^{\a} \Theta_{\a\b} \l^{\b} \Big)\, ,
\ee
where
\be\label{XITHE}
\Xi_{(\a\b)} = \Big(P_{m} +  E^{i} \bar{\t} \G_{m} \G^{\sharp} \p_{i} \t\Big) 
\G^{m}_{\a\b} +  \Pi^{m}_{i} (\G^{\sharp}\G_{m})_{\a\b} {E}^i\,,
\ee
$$
\Theta_{(\a\b)} = {1\over 2}  (\Gamma^{m}\t)_{(\a} (\Gamma^{\sharp}\Gamma_{m} \t)_{\b)} \,.
$$
Note that the second term in the r.h.s. of (\ref{nilpo}) 
vanishes due to the Gauss law $\p_{i} E^{i} =0$ constraint.

For BRST nilpotency we need to impose some conditions 
on the spinors $\l^{\a}$. In principle these conditions will depend on the form 
of the WZ term in the action via its contribution to the fermionic constraints $d_{\a}$. However, as the WZ term, being a form, is linear in time derivatives these additional pieces in $d_{\a}$ (coming from $\partial {\cal L}_{WZ} \over \partial \dot{\t}^{\a}$) are only relevant for the BRST transformations of the conjugate momenta $P_{m}$, $p_{\a}$ and $E^{i}$, but not for those of the fundamental fields $x^{m}, \t^{\a}$ and $A_{\mu}$. Hence for the latter we find the following transformation rules: 
\bea\label{trans}
&& s\, \t^{\a} = \l^{\a}\,, ~~~~~ \nn \\
&& s\, x^{m} = \bar\l \G^{m} \t\,, \nn \\
&& s\, A_{i} =  \Big[ (\bar\l \Gamma^{m } \Gamma^{\sharp} \t) \Big( \Pi_{i m} + 
{1\over 2} \bar\t \Gamma_{m} \p_{i} \t \Big) 
- {1\over 2} (\bar\l \Gamma^{m}\t) (\bar\t \Gamma_{m} \Gamma^{\sharp} \p_{i} \t) \Big] \,,
\eea
where $s$ denotes the BRST differential obtained by acting 
with $Q$ on the elementary fields.
For completness, we also list the BRST transformations of $\l^{\a}$, $\Pi^{m}_\mu$ and ${\cal F}$: 
\be\label{trasfPIF}
s\, \l^{\a} = 0\,, ~~~~~~~ s\, \Pi^{m}  = 2 \bar\l \G^{m} d \t\,, ~~~~~~~ s\, {\cal F} = 2 \bar\l \G^{m} \G^{\sharp} d \t \, \Pi_{m} \,.
\ee
Nilpotency is achieved by imposing the pure spinor constraints
\be\label{ps}
\bar\l \G^{m} \l = 0\,, ~~~~~~~~ \bar\l \G^{m} \G^{\sharp} \l =0\,.
\ee
Recall that in the Berkovits' notation they acquire the familiar form
\be \label{psS}
\l_1 \g^m \l_1 = 0 \, , ~~\qquad \l_2 \g^m \l_2 = 0 \, ,
\ee
where $\l_1$ and $\l_2$ are the two 16-component spinors that together make up the 32-component spinor $\l$ and $\g^m$ are the $16 \times 16$ blocks of the block-diagonal matrix $\cC \G^m$. These two conditions are sufficient to ensure the nilpotency of the BRST charge $Q$ 
and the BRST transformations (\ref{trans}) and 
(\ref{trasfPIF}).\footnote{Although, clearly, they are not enough for the nilpotency of the full BRST charge.} For example, 
$s^{2} \, {\cal F} = - \bar\l \G^{m} \G^{\sharp} d\l \Pi_{m} - 2 \bar\l \G^{m} \G^{\sharp} d\t \bar\l \G_{m} d\t =$ 
$- d \Big({1\over 2} \bar\l \G^{m} \G^{\sharp} \l\Big) \Pi_{m} - \bar\l \G^{m} \G^{\sharp} \l d\bar\t \G_{m} d\t = 0$ where 
the Fierz identities $(\G^{m}\G^{\sharp})_{(\a\b} \G_{m, \g) \d} =0$ have been used. 

In the case of $p$-branes the situation is simpler.\footnote{By $p$-branes we mean the superbranes of the minimal branescan (cf. the second paper in \cite{pbrane}) whose worldvolume fields comprise a scalar supermultiplet, thereby excluding not only D-branes but also the L-branes of \cite{Lbranes} (with linear supermultiplets on their worldvolume) as well as the M5-brane (with a self-dual field strength living on its worldvolume).} There is no worldvolume gauge field $A_{\mu}$ and 
therefore the BRST charge does not depend on the conjugate momentum $E^{i}$. Clearly the BRST transformation rules for $x^{m}$ and $\t^{\a}$ are the same as in (\ref{trans}). However, unlike for D$p$-branes (for which $m=0, \dots, 9$ and $\a = 1, \dots, 32$), 
for $p$-branes $m$ runs over the dimension of the allowed target space, and $\a$ is the minimal 
spinor in that dimension. The pure spinor condition $\bar\l \G^{m} \l =0$ (i.e. $\l_1 \g^m \l_1 + \l_2 \g^m \l_2 = 0$) is now sufficient to guarantee the nilpotency of (\ref{trans}) for our purposes. 

\subsection{Wess-Zumino terms from BRST cohomology} 

Both D$p$-brane and $p$-brane actions are $\kappa$-invariant due to the presence of 
a WZ term. The latter is constructed out of the superinvariants $\Pi^{m}$, $d\t^{\a}$ and ${\cal F}$ given in 
(\ref{suin}) and (\ref{gauge}) respectively. We will show that in our approach 
the WZ terms are determined by the BRST cohomology $H^{p+2}_{0}(s|d)$. 
We denote by $\Omega^{r}_{q}$ a worldvolume $q$-form with 
ghost number $r$. The cohomology $H^{p+2}_{0}(s|d)$ is computed from $H^0_{p+2}$ via the descent equations 
\be\label{desc}
s\, \Omega^{0}_{p+2} + d\, \Omega^{1}_{p+1} =0\,, 
~~~~
s\, \Omega^{1}_{p+1} + d\, \Omega^{2}_{p} =0\,,
~~~~~ \dots\,, ~~~~~~ 
s\, \Omega^{p+2}_{0} =0\,.
\ee
First we will calculate the BRST cohomology with the pure spinor constraints (\ref{ps}). 
It will turn out that the only nonvanishing groups for a (D)$p$-brane are 
$H^{k}_{p+2 -k}(s|d)$ with $k \leq 2$. Then we will show that by relaxing these constraints via the introduction of new ghosts 
the WZ term for a (D)$p$-brane becomes related to a cohomological class of $H^{p+2}_{0}(s)$. In the process we will also find the relevant extension of the BRST transformation rules. 

From the known WZ terms \cite{APS}\footnote{Although here and in the following we only use the flat space results of \cite{APS}, we should point out that the WZ terms for D-branes coupled to general supergravity backgrounds were derived in \cite{curvKappaAc}.}, the above BRST transformation rules and the descent equations (\ref{desc}) we find that the classes of $H^{2}_{p}(s|d)$ are given by selecting the proper form degree 
from the expression (the notation follows \cite{APS})
\be\label{WZ}
\Omega^{2}_{A, p} = \bar\l^{\a} e^{\cal F} C_{A,\a\b}(\Pi\!\!\!\!/)\l^{\b}\,, ~~~~~~
C_{A}(\Pi\!\!\!\!/) = \sum_{l=0} \G^{l+1}_{11} {\Pi\!\!\!\!/^{~2l} \over (2l)!}\,, ~~~~~~~{\rm for~ type~IIA}\,, 
\ee
$$\Omega^{2}_{B, p} = \bar\l^{\a} \tau_{1} e^{\cal F} C_{B,\a\b}(\Pi\!\!\!\!/)\l^{\b}\,, ~~~~~~
C_{B}(\Pi\!\!\!\!/) = \sum_{l=0} \tau^{l+1}_{3} {\Pi\!\!\!\!/^{~2l+1} \over (2l+1)!}\,, ~~~~ {\rm for~ type~ IIB}\,.
$$
Using the Fierz identities, it is straightforward to check that $s\, \Omega^{2}_{A/B, p} = 0$ because 
of the pure spinor constraints (\ref{ps}). Inverting the above logic, we  can say that the WZ terms are the lift via the descent equations (\ref{desc}) of $H^2_p$ to $H^0_{p+2}$. However, if we were to construct the WZ terms in this way, we have no reason to argue that one should start with a class of $H^2_p$ and also the structure at this level may not be unique. The situation changes dramatically in a formalism without constraints. Before turning to that, we note that the pure spinor bilinears $\bar\l \G^{\sharp} \l, \bar\l \G^{mn} \l$ etc., that appear in $\Omega_{A/B, p}^{2}$, will play an important role below.

We are finally ready to extend the 
formalism by adding new ghosts in order to avoid the constraints (\ref{ps}) on the spinors $\l^{\a}$. 
We introduce the fields $\xi^{m}$, a space-time vector with ghost number $1$, and $\Psi$ and $\Phi$, worldvolume $1$- and $0$- forms with ghost numbers $1$ and $2$ respectively. 
We also introduce the differential $\hat s = s + d$ and the following combinations
\be\label{comb}
\hat \Pi^{\a} = d \t^{\a} + \l^{\a} \,, ~~~~~~
\hat \Pi^{m} = \Pi^{m} + \xi^{m}\,, ~~~~~~
\hat {\cal F} = {\cal F} + \Psi + \Phi\,. 
\ee
The new fields have the following BRST transformation rules
\bea\label{newtrans}
\hat s \hat \Pi^{\a} = 0\,, ~~~~~ 
\hat s \hat \Pi^{m} =\hat{\overline \Pi}^{\a} \G^{m}_{\a\b} \hat \Pi^{\b} \,, ~~~~~ 
\hat s \hat {\cal F} = \hat{\overline \Pi}^{\a} (\G^{\sharp} \G^{m})_{\a\b} \hat \Pi^{\b} \hat \Pi_{m}\,.
\eea
which appear to be equivalent to the original ones but with fields and ghosts replaced by the generalized forms. The solution of the descent 
equations can now be extended to $H^{p+2}_{0}$ and thus the WZ terms are the lift of a cohomology class of the highest ghost number group. This prescription gives a unique answer as will be explained below.

It is useful to compute the BRST laws for the new ghost fields $\xi^{m}$, $\Psi$ and $\Phi$. They are
\be\label{gho}
s\, \xi^{m} = - \bar\l \G^{m} \l\,, ~~~~~ 
\ee
$$
s\, \Psi = - d \Phi - \bar\l \G^{\sharp} \G^{m} \l \Pi_{m} - 2 \bar\l \G^{\sharp} \G^{m} d \t \xi_{m}\,, 
~~~~~~~\, s\, \Phi = - \bar\l \G^{\sharp} \G^{m} \l \xi_{m}\,.
$$
These transformation rules are nilpotent without any pure spinor constraints (\ref{ps}). 
For example, acting twice on $\Phi$ we have $s^{2}\, \Phi = (\bar\l \G^{\sharp} \G^{m} \l)\, ( \bar\l \G_{m} \l)$, 
which vanishes because of the Fierz identities $\G^{11 m}_{(\a\b} \G_{m, \g)\d} =0$ for spinors 
of $SO(1,10)$ and $(\tau_{3} \G^{m})_{(\a\b} \G_{m, \g)\d} =0$ for complex Weyl spinors of $SO(1,9)$.
 
Extracting the contribution of the highest forms from (\ref{newtrans}), one obtains  
\bea\label{FDA}
d \Pi^{\a} = 0\,, ~~~~~~ d \Pi^{m} = \overline\Pi^{\a} \G^{m}_{\a\b} \Pi^{\b}\,, ~~~~~
d {\cal F} = \overline\Pi^{\a} (\G^{\sharp} \G^{m})_{\a\b} \Pi^{\b} \Pi_{m} \, ,
\eea
which are the Maurer-Cartan equations defining a free differential algebra
discussed in \cite{branealgebra, branealgebra1}. According to these papers, the WZ terms are elements 
of the Chevalley-Eilenberg (CE) cohomology of a suitably extended superspace. On the other hand, we found here that 
the WZ terms are identified with elements of the BRST cohomology of the extended BRST symmetry. The connection that equations (\ref{newtrans}) provide between these two points of view will be very important in Subsection 2.4. 

Using the above extended version of the BRST formulation (i.e. without pure spinor constraints), 
we can compute the cohomology $H^{0}_{p+2}(s|d)$ by analyzing  
the lowest groups $H^{p+2}_{0}$. The solution is
\be\label{WZnew}
\Omega_{A} = \hat{\overline \Pi}^{\a} e^{\hat{\cal F}} C_{A,\a\b} (\hat{\Pi\!\!\!\!/})\, \hat\Pi^{\b}\,, ~~~~~~
C_{A}(\hat\Pi\!\!\!\!/) = \sum_{l=0} \G^{l+1}_{11} {\hat\Pi\!\!\!\!/^{~2l} \over (2l)!}\,, ~~~~~~~{\rm for~ type~IIA}\,, 
\ee
$$\Omega_{B} = \hat{\overline\Pi}^{\a} \tau_{1} e^{\hat {\cal F}} C_{B,\a\b}(\Pi\!\!\!\!/)\, \hat\Pi^{\b}\,, ~~~~~~
C_{B}(\hat\Pi\!\!\!\!/) = \sum_{l=0} \tau^{l+1}_{3} {\hat\Pi\!\!\!\!/^{~2l+1} \over (2l+1)!}\,, ~~~~ {\rm for~ type~ IIB}\,.
$$
The highest ghost number terms in the sums (\ref{WZnew}) are 
\bea\label{DD}
&{\rm D}0&:~~~ \Omega^{2}_{A,0} = \bar\l \G^{11} \l\,, \nn \\
&{\rm D}1&:~~~ \Omega^{3}_{B,0} = \bar\l \tau_{1} \G^{m} \l \, \xi_{m} \,, \nn \\
&{\rm D}2&:~~~ \Omega^{4}_{A,0} = \bar\l \G^{11} \l \Phi + \bar\l \G^{mn} \l \, \xi_{m} \xi_{n}\,, \nn \\
&{\rm D}3&:~~~ \Omega^{5}_{B,0} = \bar\l (i \tau_{2} \G^{m}) \l \xi_{m} \Phi + \bar\l (\tau_{1} \G^{mnp}) \l \, 
\xi_{m} \xi_{n} \xi_{p} \,, 
\eea
and so on. It is easy to show that thess terms select a unique cohomological class for each D$p$-brane. For example, for 
D0 we need to construct an element with ghost number 2, which is a worldvolume scalar invariant under supersymmetry (notice that all ghosts are inert under supersymmetry transformations). In 
addition, it should also be a scalar from target space point of view. So we have only the 
combination $\Omega^{2}_{A,0} = \a \Phi + \b \bar\l \G^{11} \l$ with arbitrary coefficients $\a$, $\b$. By imposing 
BRST symmetry, one finds that $\a =0$ and $\b$ remains arbitrary. By lifting  the representative $\Omega^{2}_{A,0}$ 
to the top of the descent equations, one finds the usual WZ term $\Omega^{0}_{A,2} = d\bar\t \G^{11} d\t$. It may seem, though, that this procedure does not have a unique outcome because 
given the cohomological class $\Omega^{2}_{A,0}$, represented by the bosonic bilinear $\bar\l \G^{11} \l$, 
one can construct the powers $(\Omega^{2}_{A,0})^{n}$ for any integer $n$, which are also in the cohomology 
$H^{2}_{0}(s)$. However, they cannot be lifted to the cohomology $H^{0}_{2}$ since they are not solutions of the descent equations.\footnote{This is 
a general property. Let us consider the descent equations $s \Omega_{0} =0$, $s \Omega_{1} + d \Omega_{0} =0$ and 
$s \Omega_{2} + d \Omega_{1} =0$ as an example.  
Assuming that the first element $\Omega_{0}$ is bosonic, 
one finds a new $\Omega'_{1}$ which satisfies the second equation. In fact, $s \Omega'_{1} + d (\Omega_{0})^{n} = 0$ 
implies $s \Omega'_{1} + n (\Omega_{0})^{n-1} d \Omega_{0} =0$. Hence using $s \Omega_{1} + d \Omega_{0} = 0$, 
we have $s \Omega'_{1} - n (\Omega_{0})^{n-1} s \Omega_{1} =s \left( \Omega'_{1} - n (\Omega_{0})^{n-1} 
\Omega_{1} \right) = 0$. Therefore $\Omega'_{1} = n (\Omega_{0})^{n-1} 
\Omega_{1} + s \eta_{1}$ which implies that the representative $\Omega'_{1}$ is fermionic. Iterating this 
procedure one finds $s \Omega'_{2} + d \left( n (\Omega_{0})^{n-1} \Omega_{1} + s \eta_{1} \right) = 
s (\Omega'_{1} - d \eta_{1})  + n (n-1) (\Omega_{0})^{n-2} d \Omega_{0} \Omega_{1}$. Using again 
$s \Omega_{1} + d \Omega_{0} = 0$, we obtain $s (\Omega'_{2} - d \eta_{1}) - n (n-1) (\Omega_{0})^{n-2} 
(s \Omega_{1}) \Omega_{1} =0$. However, due to the fermionic nature of $\Omega_{1}$ 
the last term cannot be integated, namely it cannot be written as a total BRST variation.  
This means that the class $(\Omega_{0})^{n}$ cannot be lifted to the higher term in the descent equations. 
Furthermore, one can also show that the cohomological groups $H^{p}_{0}$ form a ring. Namely, given 
$\Delta_{1} \in  H^{p}_{0}$ and $\Delta_{2} \in H^{q}_{0}$, the product $\Delta_{1} \Delta_{2} $ is in the 
cohomology $H^{p+q}_0$ and can be lifted to $\Delta' $ belonging to $H^{0}_{p+q}$.}  

Another interesting example is the D$2$-brane. By acting with the BRST charge on 
$\Omega^{4}_{A,0}$ one has $s \Omega^{4}_{A,0} =  
-\bar\l \G^{11}  \bar\l \G^{11} \G^{m} \l \xi_{m} - \bar\l \G^{mn} \l \bar\l \G_{n} \l = 
-\bar\l \G^{m N}\l \bar\l \G_{N} \l = 0$ due to Fierz identities in D=11. Lifting this highest ghost number class to $H^0_{4}$ one recovers the standard D$2$-brane WZ term \cite{APS}.
 
 In order to return to the pure spinor formalism, one has to set to zero the new ghost fields $\xi_{m}, \Psi$ and 
 $\Phi $. Indeed, from their BRST transformations (\ref{gho}) it follows that
 \be\label{ghostsettozero}
\xi^{m } = 0 = - \bar\l \G^{m} \l\,, ~~~~~  s \Psi + d \Phi = 0 = - \bar\l \G^{\sharp} \G^{m} \l \Pi_{m} \,.
\ee
which give the pure spinor conditions (\ref{ps}). One may wonder whether it it correct to remove the $\Pi_m$ in the second equation of (\ref{ghostsettozero}). In favor of that we note that the truncation of the set of ghosts to a single one, $\l^{\a}$, is only consistent (meaning that BRST invariance is preserved) if the latter satisfies the pure spinor constraints as we already saw. We also point out that, in the case of superstrings, the cohomology of the extended BRST differential was shown in \cite{Grassithree} to be equivalent to the cohomology of the one defined with pure spinors. 

Now we are finally ready to add the WZ term to the action and derive the corresponding extended BRST transformation rules.\footnote{The extension refers to finding the transformation rules of all other fields. As we already explained, the BRST transformations in (\ref{trans}) and (\ref{trasfPIF}) do not depend on the WZ term.} We note that, as we will see, this leads to further constraints in the pure spinor description of D-branes. These constraints are hard to understand and, in particular, how to solve them is an issue that is currently under investigation. Despite this unresolved problem it is worth to write them down as they play a role in a relation between the present considerations and the brane superalgebras of \cite{branealgebra, branealgebra1}. We will elaborate more on that in Subsection 2.4.
 
\subsection{Extended BRST symmetry, incorporating the WZ terms}

In this subsection we discuss how the introduction of the WZ terms affects the BRST symmetry that we studied so far by re-deriving the fermionic constraints for the full ($S_{DBI}$ $+$ $S_{WZ}$) action.

Fortunately, this is easy to do using the results of \cite{hatsuda}. Namely, eq.~(\ref{ferm}) changes to
\bea\label{fermWZ}
d_{\a} &\equiv& F_{\a} = p_{\a} + P_{m} (\Gamma^{m} \t)_{\a} +  \\ 
&+& 
E^{i} \Big[ (\Gamma^{m } \Gamma^{11} \t)_{\a} ( \Pi_{i m} + {1\over 2} \t \Gamma_{m} \p_{i} \t) 
- {1\over 2} (\Gamma^{m}\t)_{\a} (\t \Gamma_{m} \Gamma^{11} \p_{i} \t) \Big] - { \p_{r} {\cal L}_{WZ} \over \p \dot{\t}^{\a}}\,. 
\nn\end{eqnarray}
Using again the ansatz $Q = \int d^{p}\sigma \bar{\l}^{\a} d_{\a}$, one finds that the BRST charge is not nilpotent but satisfies equation (\ref{nilpo}), however now with 
\bea\label{XI}
\Xi_{(\a\b)} &=& \tilde{P}\!\!\!\!/_{(\a\b)} +  \P_{i , (\a\b)} \tilde{E}^i + T_{p, (\a\b)} 
\nn \\
\Theta_{(\a\b)} &=& {1\over 2} \bar\t \Gamma^{m}_{(\a} (\bar\t \G^{\sharp} \Gamma_{m})_{\b)} \, ,
\eea
where $T_{p,\a\b}$ are $p$-forms (derived in \cite{APS}) which we have written out explicitly in Appendix B for convenience of the reader.\footnote{As is clear from (\ref{kin}), we have set the brane tension $\tau^{(p)}$ equal to one throughout the paper. If one wishes to restore it then the only change in (\ref{XI}) is that the last term in $\Xi_{(\a\b)}$ becomes $\tau^{(p)} T_{p, (\a\b)}$.} The properties of the matrices $\Xi$
are discussed in detail in \cite{hatsuda}. We only recall that $\tilde{P}_{m}$ and $\tilde{E}^{i}$ are 
the momenta $P_{m}$ and $E^{i}$ minus the contributions coming from $S_{WZ}$ whereas the last term of $\Xi$ is entirely due to the WZ term.

Clearly now the BRST charge is nilpotent if the spinor $\l^{\a}$ satisfies the constraints 
\be\label{con_WZ}
\bar\l \G^{m} \l =0\,, ~~~~~ \bar\l \G^{\sharp} \G^{m} \l =0\,,~~~~~~~
\bar\l^{\a} T^{A/B}_{p, (\a\b)} \l^{\b} = 0\,.
\ee
While the first two give the pure spinor conditions (\ref{ps}), the last one is completely new. It is a sum of several terms and nilpotency of the BRST transformations on all  fields and conjugate momenta requires the vanishing of each  term separately. More precisely the constraints for a D$p$-brane are
\bea\label{Dpure}
&&\bar\l \G^{11} \l = 0\,, ~~~~\bar\l \G^{mn} \Pi_{m, I} \l =0\,, ~~~~~~ \bar\l \G^{mnpq} (\Pi_m \wedge \Pi_n \wedge \Pi_p)_I  \, \l =0 \,, \dots ~~~~{\rm in \,\,\, IIA} \nn \\
&&\bar\l \tau_{1} \G^{m} \l = 0\,, ~~~~\bar\l (i \tau_{2}) \G^{mnp} (\Pi_m \wedge \Pi_n)_I \, \l =0\,, ~~~~~~ \dots ~~~~{\rm in \,\,\, IIB} \,,
\eea
or using (\ref{BGsg}) and the table in Appendix A:
\be \label{FC}
\l_1 \g^{m_1 ... m_k} \l_2 \, \varepsilon_{I_1...I_k} \Pi_{m_1, I_1} ... \Pi_{m_{k-1}, I_{k-1}} = 0 \, ,
\ee
\be
k = 0, 2, ... , p ~~~ {\rm in \,\,\, IIA} \qquad {\rm and} \qquad k = 1, 3, ..., p ~~~ {\rm in \,\,\, IIB} \, . \nn
\ee
The indices $I_j$ run only over the $p$ spatial directions of the $p+1$-dimensional worldvolume. Following the intuition from the string theory case, one might think that the commuting spinors are independent fields and hence the $\Pi$'s in the above constraints can be removed. However the resulting constraints are too strong, namely they seem to admit only the solution $\l_1 = \l_2 = 0$. So one is forced to keep the $\Pi$'s and thus accept that $\l_1$ and $\l_2$ depend on $x$ and $\theta$. Similar situation has already been encountered in the 
covariant study of the M$2$-brane \cite{membrane}. The supermembrane constraints that Berkovits finds are
\be \label{MC}
\bar{\l} \G^M \l = 0 \, , \qquad \bar{\l} \G^{11} \partial_I \l = 0 \qquad {\rm and} \qquad \bar{\l} \G^{MN} \l \Pi_{N, I} = 0 \, ,
\ee
where $M$ and $N$ are eleven-dimensional vector indices and $I = 1,2$. 

However even (\ref{FC}) and (\ref{MC}) seem to be too strong which raises the question: Is there a consistent nontrivial solution of the full set of constraints for a given brane and if yes, then how can one, upon utilizing it, reproduce the correct counting of bosonic and fermionic degrees of freedom? From \cite{membrane} one can see that already for the supermembrane the above questions are very hard. We do not have the full answer to them in the present context either, but we have made substantial progress in showing that the constraints contained in (\ref{FC}) are much less than the apparent number. Let us illustrate this significant reduction on the example of the D$4$-brane. For this case (\ref{FC}) gives
\be \label{Constr1}
\l_{1,\a} \l_2^{\a} = 0 \, , \qquad \l_1 \g^{mn} \Pi_{m, I} \l_2 = 0 \, , \qquad \varepsilon_{IJKL} \, \l_1 \g^{mnpq} \, \Pi_{m, I} \Pi_{n, J} \Pi_{p, K} \l_2 = 0 \, .
\ee
These constraints can be rewritten in the form:
\be \label{Constr2}
\l_{1,\a} \l_2^{\a} = 0 \, , \qquad \l_1 \g^m \P_I \l_2 = 0 \, , \qquad \varepsilon_{IJKL} \, \l_2 \g^m \, \P_I \P_J \P_K \l_1 = 0 \,
\ee
due to the following reason: In the second constraint of (\ref{Constr1}) one can use $\g^{mn} = \g^m \g^n + \eta^{mn}$ and then $\l_1 \l_2 = 0$ implies that the term with $\eta^{mn}$ drops out thus leading to the second constraint in (\ref{Constr2}). Similarly, using $\g^{mnpq} = \g^m \g^{npq} + \eta^{m [ n} \g^{pq]}$ in the third equation of (\ref{Constr1}) and dropping the $\eta$ term because of the second equation in (\ref{Constr1}) we obtain the third constraint in (\ref{Constr2}). Now, all three equations in (\ref{Constr2}) are solved by the ansatz
\be \label{ansatz}
\partial_I \l_1 = \P_I \l_2 \qquad {\rm and} \qquad \partial_L \l_2 = \varepsilon_{IJKL} \P_I \P_J \P_K \l_1 \, .
\ee
Hence the $2 \times (1+10 \times 4 + 10 \times 4)$ constraints in (\ref{Constr1}) reduce to $2 \times (11 \times 4 + 11 \times 4)$.\footnote{The reason for the overall multiplier of $2$ is that the spinors are complex. We have taken the spinor index in (\ref{ansatz}) to run over $11$ instead of $16$ values as these are the only independent ones after the pure spinor constraints (\ref{psS}) are solved.}

Now, for a D$4$-brane the above reduction may not be too drastic, but one can easily appreciate its power by realizing that, for example, for D$6$ there are $80$ more constraints in addition to the ones in (\ref{Constr1}), whereas the ansatz
\be
\partial_I \l_1 = \P_I \l_2 \qquad {\rm and} \qquad \partial_N \l_2 = \varepsilon_{IJKLMN} \P_I \P_J \P_K \P_L \P_M \l_1
\ee
contains the same number as for D$4$ and solves all relevant constraints in (\ref{FC}) in the same way as for D$4$. It is clear now how the reduction goes for all even and odd branes with dimension higher than the D$4$. For lower dimensional branes only one equation in this ansatz may be enough; namely, for D$1$ and D$2$
\be \label{Lc}
\partial_I \l_1 = \P_I \l_2
\ee
"solves" the constraints arising in (\ref{FC}). However the number of equations in (\ref{Lc}) is not smaller than the number in the original constraints so the significance of this rewriting is not clear at present.

As already mentioned, solving the above constraints and figuring out the counting of degrees of freedom is currently under investigation. So far we have understood completely only the D$0$-brane. It is rather special\footnote{Unlike for all other D-branes, for D$0$ only the first constraint in (\ref{con_WZ}) is present but not the second one, namely $\l_1 \g^m \l_1 + \l_2 \g^m \l_2 = 0$ and $\l_1 \g^m \l_1 - \l_2 \g^m \l_2 \neq 0$. This is due to the fact that $E^i = 0$ on the worldvolume.} and will be described in a separate publication. 
 
\subsection{Geometrical Interpretation}

We conclude the first part of this paper with explanation of the relation between the present analysis and recent work on D-brane geometry. More precisely, we will show that the above conditions for a nilpotent BRST charge are related to new anticommuting ghosts whose transformation properties under Lorentz symmetry imply that they belong to certain 
irreducible multiplets with the same structure as the supergroup manifold coordinates of an extended superspace studied in \cite{branealgebra, branealgebra1}. Using the relation with the CE cohomology of the latter we will derive the complete ghost specrtum associated with the full non-centrally extended algebra of \cite{branealgebra, branealgebra1}. The nilpotency of the BRST rules will follow from the closure of the algebra.

For this purpose we will use the connection with CE cohomology that equations (\ref{newtrans}), (\ref{FDA}) hinted at. In fact, it will turn out that this relation is rather deep. 
Let us first recall that, as was found in \cite{AT} for super p-branes and more recently in \cite{branealgebra1} for D-branes, the WZ terms are nontrivial elements of the CE cohomology of ordinary superspace. It was also shown \cite{branealgebra, branealgebra1} that one can extend the superalgebra with new generators such that the differentials of their dual Maurer-Cartan (MC) forms give the nontrivial cocycles. In this way the latter become coboundaries of the CE cohomology on the extended superspace and so one is able to write down a manifestly supersymmetric WZ term. The new generators are related to new group manifold coordinates which can have both vector and spinor indices. The new coordinates with spinor indices are due to the fact that the full superalgebras for (D)$p$-branes (called brane algebras) are {\it non-central} superspace extensions \cite{branealgebra, branealgebra1}. Without repeating the analysis of the above works we only mention that for a $p$-brane one needs a generator with $p$ spinorial indices.\footnote{For example, for D$2$-branes one needs the following MC forms: $\Pi$, $\Pi^m$, $\Pi^{\prime}_m$, $\Pi^{m n}$, $\Pi^{\a}$ (associated with the generator of supersymmetry $Q_{\a}$), $\Pi^{\a m}$ and $\Pi^{\a\b}$.}

We will illustrate the relation of the brane algebras with our considerations on the example of the D$4$-brane. The MC equations of the maximal central extension in this case are:
\bea\label{FDAII}
&d \Pi^{\a} = 0\,, ~~~~~~~~~
&d \Pi^{m} = \overline \Pi^{\a} \G^{m}_{\a\b} \Pi^{\b}\,, \nn \\
&d \Pi'_{m} = -  \overline \Pi^{\a} \G^{11} \G^{m}_{\a\b} \Pi^{\b}\,, ~~~~~~~~~
&d \Pi =  \overline \Pi^{\a} \G^{11}_{\a\b} \Pi^{\b} \,, \nn \\
&d \Pi^{mn} =  \overline \Pi^{\a} \G^{mn}_{\a\b} \Pi^{\b}\,, ~~~~~~
&d \Pi^{mnpq} =  \overline \Pi^{\a} \G^{mnpq}_{\a\b} \Pi^{\b}\,.
\eea
Recall from \cite{branealgebra, branealgebra1} that the extended superspace is parametrized by the coordinates $\t^{\a}$, $x^{m}$, $\phi$, $\phi^{mn}$ and $\phi^{mnpq}$. 
As in the previous section, we add for each MC form a new ghost
\bea\label{ghostsFDA}
&&\Pi^{\a} \rightarrow \Pi^{\a} + \l^{\a}\,,~~~
\Pi^{m} \rightarrow \Pi^{m} + \xi^{m}\,, ~~~
\Pi'_{m} \rightarrow \Pi'_{m} + \xi'_{m}\,, \nn \\
&&\Pi \rightarrow \Pi + \xi\,, ~~~
\Pi^{mn} \rightarrow \Pi^{mn} + \xi^{mn}\,, ~~~
\Pi^{mnpq} \rightarrow \Pi^{mnpq} + \xi^{mnpq}\,.
\eea
Now, the complete BRST transformations for the fields $\t^{\a}$, $x^{m}$, $\phi$, $\phi^{mn}$, $\phi^{mnpq}$ and ghosts $\l^{\a}$, $\xi^{m}$, $\xi'_{m}$, $\xi^{mn}$, $\xi^{mnpq}$, $\xi$ are obtained by substituting the 
generalized forms (\ref{ghostsFDA}) in (\ref{FDAII}) and replacing the differential $d$ with the 
generalized BRST differential $\hat s = s + d$. 
Indeed, one can easily verify that, for example, all the transformation rules 
for the anticommting ghosts $\xi^{m}$, $\xi'_{m}$, $\xi^{mn}$, $\xi^{mnpq}$ and $\xi$ are proportional to spinorial billinears appearing in (\ref{FC}) and thus recover the same transformations that we found before for nilpotency of the BRST charge $Q$. 

Finally, as the new coordinates of the extended group manifold of \cite{branealgebra, branealgebra1} form irreducible supermultiplets, once we have the coordinates with vector indices only we also have to include the ones with spinor (and vector) indices coming from the non-central extension of the brane superalgebra. Again, using the MC equations (cf. eqs. (21-23) in the first paper of \cite{branealgebra1}) 
and the ansatz $\hat s = s + d$ with the generalized forms $\hat \Pi_{\a} = \Pi_{\a} + \xi_{\a}$, 
$\hat \Pi^{\a m} = \Pi^{\a m} + \xi^{\a m}$ etc., we obtain the complete set 
of BRST transformation rules for the new coordinates $\phi_{\a}, \phi^{\a m}, \dots$ and ghosts. Needless to say, the above procedure works for any super (D)$p$-brane.

To summarize, the geometrical structure of the full BRST symmetry is obtained by replacing the CE cohomology of the WZ terms with the cohomology of the BRST operator $\hat s$ and the MC forms of the CE cohomology with the generalized forms given above. As the WZ terms describe the couplings of D-branes to RR backgrounds \cite{MicDouglas}, we have obtained a way to handle these couplings using BRST symmetry instead of $\kappa$-symmetry.

\section{D-branes from completely covariant open strings}
\setcounter{equation}{0}

\subsection{Preliminaries}

Consider the form of the Green-Schwarz action for a flat worldsheet given by
\be
S = \int d^2 z \left( \frac{1}{2} \partial x^m \bar{\partial} x_m + p_{\alpha} \bar{\partial} \theta^{\alpha} + \hat{p}_{\alpha} \partial \hat{\theta}^{\alpha} \right) \, , \label{GSac}
\ee
where, as in Section 2, $m$ is a space-time vector index and $\alpha$ - a space-time spinor one; in this section (unlike in the previous one) unhatted and hatted objects will denote left and right movers respectively; the partial derivatives are defined as follows
\be
\partial \equiv \partial_z = i\partial_{\tau} - \partial_{\sigma} \,\,\, ,  \qquad \bar{\partial} \equiv \partial_{\bar{z}} = i\partial_{\tau} + \partial_{\sigma} \, . \label{Der}
\ee
Note that the standard definition in (\ref{Der}) differs from the convention used in \cite{GPPN}. 

The variation of the action (\ref{GSac}) under the supersymmetry transformations\footnote{The right-hand side of (\ref{susytr}) differs by a few factors of $1/2$ from the transformation rules in \cite{GPPN} due to different definitions of the superinvariant $\Pi^m$ here and in \cite{GPPN}.}
\bea \label{susytr}
\delta_{\epsilon} \theta^{\alpha} &=& \epsilon^{\alpha} \qquad , \qquad \delta_{\hat{\epsilon}} \hat{\theta}^{\alpha} = \hat{\epsilon}^{\alpha} \nn \\
\delta_{\epsilon, \hat{\epsilon}} x^m &=& - \epsilon \gamma^m \theta - \hat{\epsilon} \gamma^m \hat{\theta} \nn \\
\delta_{\epsilon} p_{\alpha} &=& \partial x_m \gamma^m_{\alpha \beta} \epsilon^{\beta} -\frac{1}{4} \gamma^m_{\alpha \beta} \partial \theta^{\beta} (\epsilon \gamma_m \theta) \nn \\
\delta_{\hat{\epsilon}} \hat{p}_{\alpha} &=& \bar{\partial} x_m \gamma^m_{\alpha \beta} \hat{\epsilon}^{\beta} -\frac{1}{4} \gamma^m_{\alpha \beta} \bar{\partial} \hat{\theta}^{\beta} (\hat{\epsilon} \gamma_m \hat{\theta})
\eea
is
\bea
\delta_{susy} {\cal L} &=& \frac{1}{4} \left( \bar{\partial} \theta \gamma \partial \theta \, (\epsilon \gamma \theta) - \bar{\partial} \hat{\theta} \gamma \partial \hat{\theta} \, (\hat{\epsilon} \gamma \hat{\theta}) \right) \nn \\
&-& \frac{1}{2} \left( \p (\e \g \bar{\p} \t \, x) + \bar{\p} (\hat{\e} \g \p \hat{\t} \, x) \right) + \frac{1}{2} \left( \bar{\p} (\e \g \p \t \, x) + \p (\hat{\e} \g \bar{\p} \hat{\t} \, x) \right) \, . \label{Var}
\eea
Clearly the second line of (\ref{Var}) is a total derivative. One can show that this is also the case for the first line in the same way that one proves the invariance under supersymmetry of the standard GS action (see p. $\!$254 of Green, Schwarz and Witten, I). As we are interested in D-branes, we can not trow out these total derivative terms. However they vanish if we impose the following boundary conditions
\be
\theta = \pm \hat{\theta} \qquad , \qquad \epsilon = \pm \hat{\epsilon} \, . \label{TEBCs}
\ee
Indeed, for example from the first term in the second line of (\ref{Var}) we find for boundaries at fixed $\s$ and $\tau$ the contributions $\p_{\s} \left( x \, \hat{\e} \g \p \hat{\t} - x \, \e \g \bar{\p} \t \right)$ and $\p_{\tau} \left( x \, \e \g \bar{\p} \t + x \, \hat{\e} \g \p \hat{\t} \right)$ respectively. In both cases these terms cancel due to (\ref{TEBCs}) and $\p = \bar{\p} |_{\s = const}$, $\p = - \bar{\p}|_{\tau = const}$.\footnote{Of course, one should restrict to the boundary only {\it after} taking the derivative of a field. Nevertheless, saying that $\p = \bar{\p} |_{\s = const}$ and $\p = - \bar{\p}|_{\tau = const}$ is a good mnemonic rule as one can easily show that these operator relations are satisfied on every field that obeys boundary conditions of the kind of (\ref{TEBCs}) and has first order field equations.}

In the approach of \cite{GPPN} the Berkovits' pure spinor conditions are relaxed and then the covariant quantization of the superstring requires the addition of new ghost fields. The full action (only for the left-movers for brevity) becomes\footnote{We note that the field $\b_{z \a}$, which is the conjugate momentum for $\l^{\a}$, is denoted by $w_{z \a}$ in the work of Berkovits \cite{berko}.}
\be
S = \int d^2 z \left( \frac{1}{2} \partial x^m \bar{\partial} x_m + p_{z \alpha} \bar{\partial} \theta^{\alpha} + \beta_{z m} \bar{\partial} \xi^m + \beta_{z \alpha} \bar{\partial} \lambda^{\alpha} + \kappa_z^{\alpha} \bar{\partial} \chi_{\alpha} + c_z \bar{\partial} b + \omega_z^m \bar{\partial} \eta_m \right)\!. \label{CompCovAc}
\ee
The BRST variations of all fields in (\ref{CompCovAc}) are given in \cite{GPPN}. The action (\ref{CompCovAc}) is BRST invariant up to boundary terms given in (3.3) of \cite{GPPN}. At $\sigma = 0 , \pi$ the vanishing of these boundary terms can be achieved by imposing the conditions (3.5) in \cite{GPPN} which are compatible with the supersymmetry conditions (\ref{TEBCs}) with the $+$ sign. 

\subsection{Boundary conditions at fixed $\tau$}

Our primary interest in this section is in boundary conditions at fixed worldsheet time $\tau$ as we want to write down boundary states, which provide description of D-branes as coherent states of closed strings. The terms $\partial_{\tau} (...)$ in the BRST variation of the action (\ref{CompCovAc}) (supplemented with the right moving sector) vanish when the following conditions are satisfied at the boundary
\bea
\xi^m \partial_{\sigma} x_m &=& \hat{\xi}^m \partial_{\sigma} x_m \nn \\
\chi_{\alpha} \partial_{\sigma} \theta^{\alpha} &=& \hat{\chi}_{\alpha} \partial_{\sigma} \hat{\theta}^{\alpha} \nn \\
(\lambda \gamma^m \theta) \, \partial_{\sigma} x_m &=& (\hat{\lambda} \gamma^m \hat{\theta}) \, \partial_{\sigma} x_m \nn \\
(\partial_{\sigma} \theta \gamma^m \theta) (\theta \gamma_m \lambda) &=& (\partial_{\sigma} \hat{\theta} \gamma^m \hat{\theta}) (\hat{\theta} \gamma_m \hat{\lambda}) \, . \label{BT}
\eea
We will show that the above system of equations can be solved in a way compatible with the BRST transformations of all fields when one takes for $\theta$ and $\epsilon$ the standard D$p$-brane boundary conditions
\be
\theta^2 = \pm \Gamma^0 ... \,\Gamma^p \theta^1 \qquad , \qquad \epsilon^2 = \pm \Gamma^0 ... \,\Gamma^p \epsilon^1 \, , \label{BC}
\ee
where $\Gamma^m$ are the Dirac matrices, $(0, ... , p)$ are the D$p$-brane worldvolume directions and $\theta^{1,2}$ have opposite chiralities in IIA and the same chirality in IIB. In terms of the Berkovits' notation (\ref{BC}) becomes
\be
\theta = \pm \,(-1)^k \,\gamma^0 ... \,\gamma^p \,\hat{\theta} \qquad , \qquad \epsilon = \pm \,(-1)^k \,\gamma^0 ... \,\gamma^p \,\hat{\epsilon} \, , \label{BerBC}
\ee
where $p = 2k$ in type IIA and $p = 2k+1$ in type IIB. We note in passing that, as will become clear below, with (\ref{BerBC}) also the supersymmetry variation of the action (\ref{Var}) vanishes and hence the boundary conditions for $\theta$ and $\epsilon$ are determined by supersymmetry rather being imposed by hand. 

Now, as the BRST variation of $\theta$ has to be compatible with the boundary conditions:
\be
s \theta = \lambda \qquad \Rightarrow \qquad \lambda = \pm \,(-1)^k \,\gamma^0 ... \,\gamma^p \hat{\lambda} \qquad {\rm at \,\, the\,\, boundary}. \label{BCL}
\ee
Since we have to ensure that the boundary state $|{\rm B}>$ satisfies
\be
(Q + \hat{Q}) \, |{\rm B}> = 0 \, , \label{BRSTBS}
\ee
let us recall the BRST current (whose integral gives the BRST charge $Q$) derived in \cite{GPPN}\footnote{Recently a new BRST charge for the quantum superstring has been constructed in the framework of gauged WZNW models \cite{superWZW} avoiding the $b, c_{z}$ system. It should be possible to derive equivalent boundary conditions also in that formulation.}:
\bea
j^B_z &=& \lambda^{\alpha} d_{z \alpha} - \xi^m \Pi_{z m} - \chi_{\alpha} \partial_z \theta^{\alpha} - \xi^m \kappa_z^{\alpha} \gamma_{m \alpha \beta} \lambda^{\beta} -\frac{1}{2} \lambda^{\alpha} \gamma^m_{\alpha \beta} \lambda^{\beta} \beta_{z m} + c_z \nn \\
&-& \frac{1}{2} b \left( \xi^m \partial_z \xi_m - \frac{3}{2} \chi_{\alpha} \partial_z \lambda^{\alpha} + \frac{1}{2} \partial_z \chi_{\alpha} \lambda^{\alpha} \right) -\frac{1}{2} \partial_z (b \chi_{\alpha} \lambda^{\alpha}) \, . \label{cur}
\eea
Clearly (\ref{cur}) is only the holomorphic part.

From (\ref{BCL}), (\ref{BRSTBS}) and (\ref{cur}) we obtain the following boundary condition for $d_{z \alpha}$
\be
d = \mp \,(-1)^{k+p} \,\gamma^0 ... \,\gamma^p \hat{d} \, . \label{BCd}
\ee
In deriving the last equation we used the fact that
\be
(\gamma^0 ... \,\gamma^p)^t = (-1)^p \,(\gamma^0 ... \,\gamma^p)^{-1} \, . \label{ginv}
\ee
This can be proven in the following way: From (\ref{BGsg})
\be
({\cal C} \Gamma^0 ... \,\Gamma^p)^{-1} = (-1)^k \left( \begin{array}{cc} (\gamma^0 ... \,\gamma^p)^{-1} & 0 \\ 0 & (\gamma^0 ... \,\gamma^p)^{-1} \end{array} \right) \qquad {\rm in \,\,\,\, IIA} \label{A1}
\ee
and
\be
\hspace*{1.4cm}({\cal C} \Gamma^0 ... \,\Gamma^p)^{-1} = (-1)^k \left( \begin{array}{cc} 0 & - (\gamma^0 ... \,\gamma^p)^{-1} \\ (\gamma^0 ... \,\gamma^p)^{-1} & 0 \end{array} \right) \hspace*{0.5cm} {\rm in \,\,\,\, IIB} \, . \label{B1}
\ee
On the other hand
\be
(\Gamma^0 ... \,\Gamma^p)^{-1} = (-1)^p \,{\cal C}^{-1} \,(\Gamma^0 ... \,\Gamma^p)^t \,{\cal C}
\ee
and hence
\be
({\cal C} \Gamma^0 ... \,\Gamma^p)^{-1} = (\Gamma^0 ... \,\Gamma^p)^{-1} {\cal C}^{-1} = (-1)^p {\cal C}^{-1} (\Gamma^0 ... \,\Gamma^p)^t = (-1)^p {\cal C}^{-1} ({\cal C} \Gamma^0 ... \,\Gamma^p)^t \,{\cal C} \, . \label{Interms}
\ee
In the last equality we have used that ${\cal C}^t = {\cal C}^{-1}$. Now (\ref{Interms}), together with (\ref{BGsg}) and (\ref{ChargeConj}), gives
\be
({\cal C} \Gamma^0 ... \,\Gamma^p)^{-1} = (-1)^{p+k} \left( \begin{array}{cc} (\gamma^0 ... \,\gamma^p)^t & 0 \\ 0 & (\gamma^0 ... \,\gamma^p)^t \end{array} \right) \qquad {\rm in \,\,\,\, IIA} \label{A2}
\ee
and
\be
\hspace*{1.4cm}({\cal C} \Gamma^0 ... \,\Gamma^p)^{-1} = (-1)^{p+k} \left( \begin{array}{cc} 0 & - (\gamma^0 ... \,\gamma^p)^t \\ (\gamma^0 ... \,\gamma^p)^t & 0 \end{array} \right) \hspace*{0.5cm} {\rm in \,\,\,\, IIB} \, . \label{B2}
\ee
Comparing (\ref{A1}) with (\ref{A2}) and (\ref{B1}) with (\ref{B2}) we obtain in all cases (\ref{ginv}).

Now we turn to the boundary condition for $\beta_{z m}$. The fifth term in (\ref{cur}) implies that it is determined by the boundary condition for $\lambda$. Crucial are also the relations
\bea
&&(\gamma^0 ... \,\gamma^p)^t \,\, \gamma^m \,\, (\gamma^0 ... \,\gamma^p) = \gamma^m  \,\,\, , \qquad m\in (0,...,p) \hspace*{0.5cm} {\rm i.e.} \hspace*{0.4cm} {\rm N} \nn \\
&&(\gamma^0 ... \,\gamma^p)^t \,\, \gamma^m \,\, (\gamma^0 ... \,\gamma^p) = - \gamma^m \,\,\, , \qquad m\in\!\!\!\!\!/ \,\, (0,...,p) \hspace*{0.5cm} {\rm i.e.} \hspace*{0.4cm} {\rm D} \, , \label{Rel}
\eea
where N denotes Neumann and D - Dirichlet direction. These relations follow easily from
\bea
&&\hat{{\rm P}} \, \Gamma^m \, {\rm P} = \Gamma^m \,\,\, , \qquad m\in (0,...,p) \nn \\
&&\hat{{\rm P}} \, \Gamma^m \, {\rm P} = - \Gamma^m \,\,\, , \qquad m\in\!\!\!\!\!/ \,\, (0,...,p) \, ,
\eea
where
\be
{\rm P} = \Gamma^0 ... \,\Gamma^p \qquad {\rm and} \qquad \hat{{\rm P}} = {\cal C}^{-1} \, {\rm P}^t \, {\cal C} \, .
\ee
Indeed
\be
\Gamma^m = \pm \, \hat{{\rm P}} \, \Gamma^m \, {\rm P} = \pm \, \cC^{-1} \, {\rm P}^t \, (\cC \Gamma^m) \, {\rm P} \, .
\ee
Multiplying from the left by $\cC$ and inserting $\cC \cC^{-1}$ in two places we obtain
\be
(\cC \Gamma^m) = \mp \, (\cC {\rm P})^t \, \cC^{-1} \, (\cC \Gamma^m) \, \cC^{-1} \, (\cC {\rm P}) \, .
\ee
This equation, combined with (\ref{BGsg}), gives for both $p$ even and odd the relations (\ref{Rel}). Using the latter and (\ref{BCL}) we find
\be
{\rm N} \, : \,\,\,\, \beta_{z m} = - \hat{\beta}_{z m} \qquad ; \qquad {\rm D} \, : \,\,\,\, \beta_{z m} = \hat{\beta}_{\bar{z} m} \, . \label{BzBC}
\ee

Now we turn to the second term in (\ref{cur}) which will allow us to determine the boundary conditions for $\xi^m$ from the ones for $\Pi_{z m}$. Recall from (\ref{suin}) that 
\be
\Pi^m = \partial x^m + \theta^{\alpha} \gamma^m_{\alpha \beta} \partial \theta^{\beta} \qquad {\rm and} \qquad \hat{\Pi}^m = \bar{\partial} x^m + \hat{\theta}^{\alpha} \gamma^m_{\alpha \beta} \bar{\partial} \hat{\theta}^{\beta} \, . \label{PiC}
\ee
As we are interested in a boundary at fixed $\tau$ the Neumann (worldvolume) directions $m \in (0, ..., p)$ are the ones for which $\partial_{\tau} x^m = 0$. In the Dirichlet (transverse) directions $m\in\!\!\!\!\!\!/ \,\, (0,...,p)$ the boundary condition for $x$ is $x^m = const$ meaning in particular that $\partial_{\sigma} x^m = 0$. Hence from (\ref{Der}), (\ref{BerBC}) and (\ref{Rel}) (together with the fact that $\partial = - \bar{\partial} \, |_{\tau \,= \,const}$ in the sense explained in Subsection 3.1.) we obtain
\be
{\rm N} \, : \,\,\,\, \Pi = - \hat{\Pi} \qquad ; \qquad {\rm D} \, : \,\,\,\, \Pi = \hat{\Pi} \, . \label{PiBC}
\ee
Note that this is opposite to the more familiar case when the worldsheet boundary is at fixed $\sigma$. In that case
\be
{\rm N} \, : \,\, \partial_{\sigma} x^m = 0 \qquad ; \qquad {\rm D} \, : \,\, x^m = const \qquad ; \qquad \partial = \bar{\partial} \,|_{\sigma \,= \,const} \nn 
\ee
\be
\Longrightarrow \hspace*{1.2cm} {\rm N} \, : \,\, \Pi = \hat{\Pi} \qquad ; \qquad {\rm D} \, : \,\, \Pi = - \hat{\Pi} \, . \label{BCPi}
\ee
From (\ref{PiBC}), (\ref{BRSTBS}) and the second term in (\ref{cur}) we obtain
\be
{\rm N} \, : \,\,\,\, \xi^m = \hat{\xi}^m \qquad ; \qquad {\rm D} \, : \,\,\,\, \xi^m = - \hat{\xi}^{m} \, . \label{XiBC}
\ee

So far we have determined the boundary conditions of all but two fields in the first line of the BRST current: $c_z$ and $\kappa^{\alpha}_z$. The first one is obvious:
\be
c = - \hat{c} \qquad {\rm at \,\, the \,\, boundary} \, .
\ee
The second can be found from the fourth term of (\ref{cur}) upon using (\ref{Rel}). It is
\be
\kappa = \mp \,(-1)^{k} \,\gamma^0 ... \,\gamma^p \hat{\kappa} \, .
\ee

At this point the only fields in $j^B$, that remain to be treated, are $\chi_{\alpha}$ and $b$ both entering (\ref{cur}) in terms containing derivatives. Using once more that on the boundary $\partial = - \bar{\partial}$ together with (\ref{ginv}) and the already known conditions for $\lambda$ and $\xi^m$ we find
\be \label{chib}
\chi_{\alpha} = \pm \,(-1)^{k+p} \,(\gamma^0 ... \,\gamma^p \hat{\chi})_{\alpha} \qquad {\rm and} \qquad b = \hat{b} \,\, .
\ee

However we still have to determine the boundary condition for $\beta_{z \alpha}$. For that purpose let us make the following observation: The boundary conditions that we found above for $\xi^m$, $\beta_{z m}$, $\kappa_z^{\alpha}$, $\chi_{\alpha}$, $b$ and $c_z$, together with (\ref{ginv}), imply that
\be
\beta_{z m} \xi^m = - \hat{\beta}_{\bar{z} m} \hat{\xi}^m \, , \qquad \kappa_z^{\alpha} \chi_{\alpha} = - \hat{\kappa}_{\bar z}^{\alpha} \hat{\chi}_{\alpha} \, , \qquad b c_z = - \hat{b} \hat{c}_{\bar{z}} . \label{Prod}
\ee
Fortunately these products are terms in the ghost current \cite{GPPN}:
\be
J_z^{gh} = - (\beta_{z m} \xi^m + \kappa_z^{\alpha} \chi_{\alpha} + \beta_{z \alpha} \lambda^{\alpha} + b c_z + \eta^m \omega_{z m}) \, .
\ee
So (\ref{Prod}) suggests that at the boundary\footnote{In fact, this is also the boundary condition for the ghost current $J^{gh} = b c + \b \g$ in the NSR formalism, where at a fixed $\tau$ boundary $c = - \hat{c}$, $b = \hat{b}$, $\b = \pm i \hat{\b}$ and $\g = \pm i \hat{\g}$ \cite{CNLY1, CNLY2}. In addition, one can easily check that the same boundary condition is satisfied by the grading current $J^{gr}_{z}$ introduced in \cite{grassi} to select physical states.}
\be
J_z^{gh} = - \hat{J}_{\bar{z}}^{gh} \, . \label{ghBC}
\ee
From the last equation, the boundary condition for $\lambda$ and relation (\ref{ginv}) it follows that
\be
\beta_{\alpha} = \mp \,(-1)^{k+p} \,(\gamma^0 ... \,\gamma^p \hat{\beta})_{\alpha} \qquad {\rm at \,\, the \,\, boundary} \,\, . \label{sevBC}
\ee

Finally, we turn to the conjugate pair $\omega_{z m}$, $\eta^m$. As these fields are inert under both supersymmetry and BRST symmetry we are free to choose the boundary conditions for one of them and then the conditions for the other follow from (\ref{ghBC}). Guided by the index structure, we take the conditions for $\omega_{z m}$ to be the same as the ones for $\b_{z m}$. Hence
\be
{\rm N} \, : \,\,\,\, \omega_{z m} = - \hat{\omega}_{z m} \,\,\,\, ,  \,\,\,\, \eta^m = \hat{\eta}^m \qquad ; \qquad {\rm D} \, : \,\,\,\, \omega_{z m} = \hat{\omega}_{z m} \,\,\,\, ,  \,\,\,\, \eta^m = - \hat{\eta}^{m} \, .
\ee

Now we have all boundary conditions. We have checked that they are compatible with the BRST variations of all fields as derived in \cite{GPPN} and that also equations (\ref{BT}) are satisfied.\footnote{In these calculations one often makes use of (\ref{ginv}) and (\ref{Rel}).} Recall that (\ref{BT}) are the conditions for the BRST invariance of the superstring action when boundaries (D-branes) are present. Instead of boring the reader with all the details we will only illustrate the subtle interplay between the various boundary conditions and properties of $\gamma^m$'s on the examples of the first and third equations in (\ref{BT}):
\bea
\xi^m \partial_{\sigma} x_m &=& \hat{\xi}^m \partial_{\sigma} x_m \label{Eq1} \\
(\lambda \gamma^m \theta) \partial_{\sigma} x_m &=& (\hat{\lambda} \gamma^m \hat{\theta}) \partial_{\sigma} x_m \, . \label{Eq2}
\eea
Indeed these equations may need some explanation as one might think that we have just forgotten to cancel $\partial_{\sigma} x_m$ on both sides of each of them. In fact the presence of this multiplier turns out to be absolutely crucial: Looking at the boundary conditions for $\xi^m$ (\ref{XiBC}) we see that for $m\in$ N (Neumann directions) things are O.K., but for $m\in$ D there is a clash with (\ref{Eq1}). What saves the day is that $\partial_{\sigma} x_m = 0$ for a Dirchlet direction as explained below (\ref{PiC}). Similarly (\ref{Rel}) implies that (\ref{Eq2}) is satisfied for $m\in$ N, whereas for $m\in$ D there would have been a contradiction if both sides were not multiplied by $\partial_{\sigma} x_m = 0$.

\subsection{Boundary conditions at fixed $\sigma$}

Before writting the operator solutions of the above boundary conditions which give the boundary states describing the emission or absorbtion of closed strings from a D$p$-brane, let us briefly consider the case of boundaries at fixed worldsheet space $\sigma$ (usually at $\sigma = 0, \pi$). The analysis is very similar to the one in Subsection 3.2. The boundary terms coming from supersymmetry variation of the action are canceled for the same conditions (\ref{BerBC}) for $\theta$ and $\epsilon$. This and BRST invariance imply again (\ref{BCL}) for $\lambda$ which in turn leads again to (\ref{BCd}) and (\ref{BzBC}) for $d$ and $\b_{z m}$ respectively. The difference starts with $\Pi^m$, the boundary conditions for which change sign as explained in (\ref{BCPi}). From this it follows that for $\xi^m$ there is also a change of sign
\be
{\rm N} \, : \,\,\,\, \xi^m = - \hat{\xi}^m \qquad ; \qquad {\rm D} \, : \,\,\,\, \xi^m =  \hat{\xi}^{m} \,\, , \label{newxiBC}
\ee
which implies the corresponding change for $\kappa$:
\be
\kappa = \pm \,(-1)^{k} \,\gamma^0 ... \,\gamma^p \hat{\kappa} \,\, .
\ee
Now\footnote{We note that again the condition on the ghost current coincides with the one in the NSR formalism.}
\be
\b_{z m} \xi^m = \hat{\b}_{\bar{z} m} \hat{\xi}^m \qquad \Longrightarrow \qquad J^{gh}_z = \hat{J}^{gh}_{\bar{z}} \hspace*{0.5cm} {\rm at \,\, the \,\, boundary} \,\, .
\ee
Hence the sign of the $\b_{\a}$ boundary condition also changes:
\be
\beta_{\alpha} = \pm \,(-1)^{k+p} \,(\gamma^0 ... \,\gamma^p \hat{\beta})_{\alpha}
\ee
whereas the one for $\chi_{\a}$ remains the same as in (\ref{chib}). Similarly to the logic in the previous subsection, the boundary conditions for $\eta^m$ become the same as (\ref{newxiBC}). Finally, as nothing changes for $c$ the new boundary condition for the ghost current implies\footnote{Clearly one can determine the conditions for $\chi_{\alpha}$ and $b$ as in the previous subsection, namely directly from the BRST current using the fact that now $\partial = \bar{\partial}$ at the boundary. As is to be expected, both ways give the same answer.}
\be
b = - \hat{b} \, .
\ee

One can show that with the above bundary conditions the boundary terms coming from BRST variation of the superstring action vanish. We only note that the difference with equations (\ref{BT}), which guaranteed this vanishing for boundary at fixed $\tau$, is that $\partial_{\sigma}$ now becomes $\partial_{\tau}$. This change is exactly what is needed to compensate the change of signs of other boundary conditions described in this subsection.

Hence we have proven the BRST invariance of the action of open strings ending at $\sigma = 0, \pi$ on D$p$-branes. As is well-known this invariance is the analog of the kappa symmetry of the Green-Schwarz action. In \cite{LW} the latter was used to infer that in flat space the D-branes of type IIA string theory must have odd dimensional worldvolume whereas those of type IIB - even-dimensional one. The argument is the following: varing the GS action under kappa symmetry one finds boundary terms which can be canceled by imposing (\ref{BC}). These boundary conditions together with the fact that $\theta^1$ and $\theta^2$ have opposite chiralities in IIA and the same chirality in IIB imply the above conclusion about the dimensionality of the D-brane worldvolumes in the two theories. As we have seen, the same boundary conditions are compatible with BRST invariance and hence the above conclusion can also be viewed as a consequence of the latter symmetry.

\subsection{Boundary states}

We finally turn to the operator solution of the boundary conditions at fixed $\tau$ found in Subsection 3.2. Let us first mention that one can easily recover from them the conditions for free open strings in the following way: As an open string with free end points can be described as being constrained to end on a D9-brane, we find using (\ref{g10}) that the conditions (\ref{BerBC}) become
\be
\theta = \pm \hat{\theta} \qquad , \qquad \epsilon = \pm \hat{\epsilon} \, .
\ee
Similarly, all other boundary conditions reduce to the usual ones written, for example, in \cite{GPPN, BP}.

Now, as T-duality exchanges Neumann and Dirichlet boundary conditions one obtains the lower dimensional D-branes from the D9-brane by T-dualising along an appropriate number of worldvolume directions. As under T-duality the only change for the oscilators of the NSR formalism is in the sign of the right-movers, the boundary state for a D$p$-brane is obtained from the boundary state of \cite{CNLY1, CNLY2}, describing the creation of a closed string from the vaccum, by changing the sign of the terms corresponding to Dirichlet directions. To illustrate this, recall that the boundary condition $\partial_{\tau} x^m = 0$ gives the following relations between the left and right moving closed string modes\footnote{We use the conventions of \cite{CNLY1} with the only difference that, as in the rest of this section, we denote the right-movers by \,$\hat{}$ \,instead of \,\,$\tilde{}$\,.}
\be
a_n^{\dagger} e^{n \tau} = \hat{a}_n e^{-n \tau} \, , \qquad \hat{a}_n^{\dagger} e^{n \tau} = a_n e^{-n \tau} \, , \qquad n \ge 1.
\ee
These conditions are solved in terms of coherent states \cite{CNLY1}:
\be
|{\rm B}_x> = \exp \left( \sum_{n=1}^{\infty} e^{2n \tau} (a_n^{0 \, \dagger} \hat{a}_n^{0 \, \dagger} - a_n^{i \, \dagger} \hat{a}_n^{i \, \dagger}) \right) |0> \qquad {\rm where} \qquad i=1,...,9 \, .
\ee
Taking as an example a D2-brane along $x^0, x^1, x^2$\,, we find that the above expression becomes
\be
\hspace*{-0.4cm} |{\rm B}_x> = \exp \!\left( \sum_{n=1}^{\infty} e^{2n \tau} (a_n^{0 \, \dagger} \hat{a}_n^{0 \, \dagger} - a_n^{1 \, \dagger} \hat{a}_n^{1 \, \dagger} - a_n^{2 \, \dagger} \hat{a}_n^{2 \, \dagger} + a_n^{j \, \dagger} \hat{a}_n^{j \, \dagger}) \!\right) \!|0> \hspace*{0.2cm} , \hspace*{0.4cm} j=3,...,9 \, .
\ee

The full boundary state in the NSR formalism is
\be
|{\rm B}> = |{\rm B}_x> |{\rm B}_{\psi}> |{\rm B}_{gh}> \, .
\ee
In the BRST formulation of \cite{GPPN} the differences with the last formula are of two kinds. One is simply that the set of ghosts (let us denote it by $\tilde{gh}$) is different from the $b,c$ and $\beta, \gamma$ ghosts of the NSR string. And the other is in the fact that now there are fields with spacetime spinor index. Hence the boundary state for a D$p$-brane is
\be
|{\rm B}> = |{\rm B}_x> |{\rm B}_{\theta}> |{\rm B}_{\tilde{gh}}> \, ,
\ee
where $|{\rm B}_{\theta}>$ is determined from the boundary condition (\ref{BerBC}) to be
\be
|{\rm B}_{\theta}> = \exp \left( \pm (-1)^k \sum_{n=1}^{\infty} e^{2n \tau} (\theta_n^{\dagger} \gamma^{0 ... p} \hat{\theta}_n^{\dagger}) \right) \!|0> \, . \label{BST}
\ee
In (\ref{BST}) $\theta_n^{\dagger}$ are the closed string creation modes in the expansion of $\theta$. We remind the reader that the $+/-$ sign corresponds to a brane/anti-brane respectively. In a similar way one can determine the boundary state contributions of all ghosts from their boundary conditions derived in Subsection 3.2.

As the covariant formulation of \cite{GPPN} is valid for flat space only, writing boundary states in it may seem a rather modest achievement. However, recall that this is not possible in the Berkovits' formulation as in the latter some fields (namely the commuting spinor $\lambda$) are constrained unlike in \cite{GPPN} where all fields are free. We regard the current section as only a first step towards the study of D-brane boundary states in nontrivial backgrounds in a {\it completely covariant formulation}. Of course such a study goes through further developing the formalism of \cite{GPPN} for arbitrary backgrounds.

\section{Conclusions and outlook}

It is strongly believed that the BRST formulation of superstrings \cite{berko} and \cite{GPPN, grassi} is 
  a viable alternative to the usual RNS and GS formalisms. In this paper we found further evidence in support of such a claim by showing that BRST symmetry in a formulation without pure spinors provides a derivation (alternative to the $\kappa$-symmetric one) of the Wess-Zumino terms in the effective actions for D$p$-branes. We also studied D-branes from the point of view of completely covariant open strings: We found the boundary conditions for all fields in the formalism of \cite{GPPN} and wrote down the corresponding boundary state. This is an initial step towards the computation of the one loop open string diagram describing the interaction between two branes at lowest order in the string coupling. Completing such a calculation would have to await understanding how are diffeomorphisms encoded in the BRST formulation. We hope to come back to this issue in the future.

Recently one of the present authors and collaborators showed that the formalism of \cite{GPPN, grassi} for covariant description of the superstring in flat 10-dimensional space-time is equivalent to a WZNW model based on an extended super-Poincar\'e algebra 
\cite{superWZW}. On the other hand, 
D-branes in WZNW models have been discussed extensively in the literature 
\cite{DbraneWZW}. It is therefore pressing to apply this well-developed machinery to the construction of \cite{superWZW} in order to see how are D-branes encoded in the latter. Furthermore, as the new formalism is target space supersymmetric its generalization to curved space would open the door for the study of generic Ramond--Ramond backgrounds which are attracting a great deal of attention at present (see, for example, \cite{backDbranes} for investigation of D-branes in $AdS_{5} \times S^{5}$ and pp-wave backgrounds). 
This subject will be studied in future publications. 

In conclusion, let us also mention that there are still a number of open issues. To list a few of them: {\it i)} What is the role 
of diffeomorphisms and gauge symmetries in the present framework? {\it ii)} How does the counting of degrees of freedom for (D)$p$-branes work? {\it 
iii)} Can one study the spectrum of quantized (D)$p$-branes? 
{\it iv)} Can one study interactions between different types of branes and 
intersecting branes? 

\section*{Acknowledgements} 

We have benefited from discussions with P. van Nieuwenhuizen and 
W. Siegel. P.A.G. thanks the Theory Division at CERN for its hospitality during the completion of this work. This research was partly supported by NSF grant PHY-0098527.

\appendix
\section{Properties of $\Gamma$ matrices}

In any number $D$ of space-time dimensions the matrices ${\cal C} \Gamma^{m_1...m_p}$ with $p < D$, where ${\cal C}$ is the charge conjugation matrix and $\Gamma^{m_1...m_p}$ is the antisymmetrized product of Dirac matrices, have definite symmetry property (determined by $p$) w.r.t. interchange of the two spinor indices. This is due to
\be
\Gamma^{m \,t} = - {\cal C} \Gamma^m {\cal C}^{-1} \qquad {\rm and} \qquad {\cal C} = \pm {\cal C}^t \, , \label{GC}
\ee
where the choice of sign in the second equation is determined by $D$. In {\it ten} dimensions
\be
{\cal C} = - {\cal C}^t \, . \label{CC}
\ee
It is easy to see that this implies that ${\cal C} \Gamma^{mn}$ is a symmetric matrix, ${\cal C} \Gamma^{mns}$ is antisymmetric etc. as shown in the table below. Indeed, let us consider the following product:
\be
(\Gamma^{m_p} ... \,\Gamma^{m_1})^t = \Gamma^{m_1 \,t} ... \,\Gamma^{m_p \,t} = (-1)^p \,{\cal C} \Gamma^{m_1} ... \,\Gamma^{m_p} {\cal C}^{-1}\, ,
\ee
where the last equality is due to the first equation in (\ref{GC}). Multiplying from the right by ${\cal C}$ and using (\ref{CC}) in the left-hand side of the equation, we obtain
\be
({\cal C} \Gamma^{m_p} ... \,\Gamma^{m_1})^t = (-1)^{p+1} \,{\cal C} \Gamma^{m_1} ...\,\Gamma^{m_p} = (-1)^{p+1 + \frac{p(p-1)}{2}} \,{\cal C} \Gamma^{m_p} ... \,\Gamma^{m_1} + \sum g^{m_i \,m_j} \,\Gamma^{...} \, .
\ee
The sum containing $g^{m_i \,m_j}$ comes from anticommuting the gamma matrices and we did not write it explicitly as it will drop out in a moment. Namely, antisymmetrizing the last equation w.r.t. the indices $m_1, ... ,m_p$ we find
\be
({\cal C} \Gamma^{m_1 ... m_p})^t = (-1)^{p+1 + \frac{p(p-1)}{2}} \,{\cal C} \Gamma^{m_1 ... m_p} \, ,
\ee
which implies the symmetry properties recorded on the second row of the table (clearly $s$ stands for symmetric and $a$ - for antisymmetric).

\vspace{0.5cm}

\begin{center}
\begin{tabular}{|c|c|c|c|c|c|c|c|c|c|c|} \hline $p$ & 1 & 2 & 3 & 4 & 5 & 6 & 7 & 8 & 9 & 10 \\ \hline  ${\cal C} \Gamma^{m_1 ... m_p}$ & $s$ & $s$ & $a$ & $a$ & $s$ & $s$ & $a$ & $a$ & $s$ & $s$ \\ \hline $\gamma^{m_1 ... m_p} \, , \,\, m_i \neq 0 \,\, \forall i$ & $s$ & $a$ & $a$ & $s$ & $s$ & $a$ & $a$ & $s$ & $s$ & $a$ \\ \hline  $\gamma^{0 m_2 ... m_p}$ & $s$ & $s$ & $a$ & $a$ & $s$ & $s$ & $a$ & $a$ & $s$ & $s$ \\ \hline \end{tabular}
\end{center}

\vspace{0.5cm}

Berkovits uses the following convenient notation
\be
{\cal C} \Gamma^m = \left( \begin{array}{cc} \gamma^m & 0 \\ 0 & \gamma^m \end{array} \right) \, . \label{Berkg}
\ee
Clearly $({\cal C} \Gamma^m)_{\alpha \beta} = ({\cal C} \Gamma^m)_{\beta \alpha}$ implies that $\gamma^m_{\alpha \beta} = \gamma^m_{\beta \alpha}$. We will need the (anti)symmetry properties w.r.t. spinor indices of the antisymmetrized products $\gamma^{m_1 ... m_p}$. Let us start with two gamma matrices:
\be
{\cal C} \Gamma^m \Gamma^n = ({\cal C} \Gamma^m) \,{\cal C}^{-1} \,({\cal C} \Gamma^n) = \left( \begin{array}{cc} 0 & - \gamma^m \gamma^n \\ \gamma^m \gamma^n & 0 \end{array} \right) \, , \label{twoG}
\ee
where we have used for ${\cal C}^{-1}$ the explicit inverse of the charge conjugation matrix
\be
{\cal C} = \left( \begin{array}{cc} 0 & 1_{16\times 16} \\ -1_{16\times 16} & 0 \end{array} \right) \, . \label{ChargeConj}
\ee
Antisymmetrizing (\ref{twoG}) w.r.t. $m$ and $n$ we obtain
\be
{\cal C} \Gamma^{mn} = \left( \begin{array}{cc} 0 & - \gamma^{m n} \\ \gamma^{m n} & 0 \end{array} \right) \, ,
\ee
which gives $\gamma^{m n}_{\alpha \beta} = - \gamma^{m n}_{\beta \alpha}$. Applying the same idea, namely inserting ${\cal C}^{-1} \,{\cal C}$ between every two successive $\Gamma^m$'s, we find that
\bea
{\cal C} \Gamma^{m_1 ... m_{2k+1}} &=& (-1)^k \left( \begin{array}{cc} \gamma^{m_1 ... m_{2k+1}} & 0 \\ 0 & \gamma^{m_1 ... m_{2k+1}} \end{array} \right) \, , \qquad k = 0, ... ,4 \nn \\ \nn \\
{\cal C} \Gamma^{m_1 ... m_{2k}} &=& (-1)^k \left( \begin{array}{cc} 0 & \gamma^{m_1 ... m_{2k}} \\ - \gamma^{m_1 ... m_{2k}} & 0 \end{array} \right) \, , \qquad k = 1, ... ,5 \, . \label{BGsg}
\eea
The symmetry properties of $\gamma^{m_1 ... m_p}$ that follow from (\ref{BGsg}) are recorded on the third row of the above table. 

In fact, these considerations are correct only when $p$ is odd or when it is even but no spacetime index of $\gamma^{m_1 ... \,m_p}$ is zero. The reason is that $\gamma^m_{\a \b}$ and $\gamma^{m \a \b}$ are {\it not} equal numerically as the Berkovits' notation (\ref{Berkg}) misleadingly suggests. Indeed, using the gamma-matrix representation in appendix A of \cite{GPN} one can see that numerically $\gamma^m_{\a \b} = - \gamma^{m \a \b}$ for $m \neq 0$ and $\gamma^0_{\a \b} = \gamma^{0 \a \b}$. As a result the matrices $\gamma^{m_1 ... \,m_p}$ on the upper and lower rows of $\cC \Gamma^{m_1 ... \,m_p}$ in (\ref{BGsg}) are only equal up to a sign. This clearly does not affect the (anti)symmetry properties of $\gamma^{m_1 ... \,m_{2k+1}}$ and furthermore as products of odd number of $\gamma$'s appear in type IIB where spinors have the same chirality the different relative sign between the upper left and lower right corners in $\cC \Gamma^{m_1 ... \,m_{2k+1}}$ has no consequence. However as the matrix $\cC \Gamma^{m_1 ... \,m_{2k}}$ is block off diagonal the (anti)symmetry properties of its blocks do depend on the relative sign between them and also such products appear in type IIA where there are spinors of both chiralities so this relative sign has additional implications. To be more precise let us denote
\be
\cC \Gamma^{m_1 ... m_{2k}} = (-1)^k \left( \begin{array}{cc} 0 & \gamma^{m_1 ... m_{2k}} \\ - \tilde{\gamma}^{m_1 ... m_{2k}} & 0 \end{array} \right) \, , \qquad k \in \bb{Z}_{\bf +} \,\, .
\ee
Making use of the representation in appendix A of \cite{GPN} one finds that 
\be
\g^{m_1 ... \,m_{2k}} = \tilde{\g}^{m_1 ... \,m_{2k}} \, , \,\,\,\, m_i \neq 0 \,\,\,\, \forall i \qquad {\rm and} \qquad \g^{0 \,m_2 ... \,m_{2k}} = - \tilde{\g}^{0 \,m_2 ... \,m_{2k}} \, . \label{g0m}
\ee
Hence we conclude that the (anti)symmetry properties of $\g^{0 \,m_2 ... \,m_p}$ for $p = 2k$ are opposite w.r.t. to the case of all spacetime indices being different from zero as on the fourth row of the above table.

Before closing this appendix we also note that
\be {\cal C} \Gamma^{m_1 ... m_{10}} = {\cal C} \Gamma^{11} = - \left( \begin{array}{cc} 0 & 1_{16\times 16} \\ 1_{16\times 16} & 0 \end{array} \right) \, ,
\ee
which, in particular, implies
\be
\g^{0...9} = 1_{16 \times 16} \qquad {\rm and} \qquad \tilde{\g}^{0...9} = - 1_{16 \times 16} \, . \label{g10}
\ee

\section{WZ terms from $\kappa$-invariance}

The effective world-volume D$p$-brane action is
\be
S = - \int d^{p+1} \sigma \sqrt{-\det (G_{\mu \nu} + {\cal F}_{\mu \nu})} \,\,\, + \,\,\, S_{WZ} \,  \label{action}
\ee
with $G_{\mu \nu}$ and ${\cal F}_{\mu \nu}$ as in Subsection 2.1. The term $S_{WZ} = \int d^{p+1}\sigma \, L_{WZ}$ was found in \cite{APS}\footnote{As mentioned in the text, the WZ terms for D-branes coupled to general backgrounds were found in \cite{curvKappaAc} by requiring $\kappa$-invariance. However, as such backgrounds are beyond the scope of this paper we are only using the simpler expressions of \cite{APS}.} from the DBI action by imposing the requirement that the full action $S$ be invariant under $\kappa$-symmetry. The result is that
\be\label{dWZ}
d L^{WZ}_{A/B} = d\bar \t \, T_{A/B} \, d\t \,,
\ee
where for each D$p$-brane one has to take the appropriate term in the formal sum of differential forms
\be
T_A = \sum_{p \,- \,{\rm even}} T_p \, , \qquad T_B = \sum_{p \,- \,{\rm odd}} T_p
\ee
for IIA and IIB string theory respectively. The forms $T_p$ are given by
\bea
&&\hspace*{-5cm}T_{0} = \Gamma^{11} \nn\\ \nn \\
&&\hspace*{-5cm}T_{1} = \tau_{1} \P \nn 
\eea
\bea
&&\hspace*{-5cm}T_{2} = \frac{\P^{\, 2}}{2!} + {\cal F} \Gamma^{11}   \nn \\ \nn \\
&&\hspace*{-5cm}T_{3} = i \tau_2 \frac{\P^{\, 3}}{3!} + {\cal F} \P \tau_1 \nn \\ \nn 
\eea
\bea
&&\hspace*{-4.6cm}T_{4} = \Gamma^{11}
\frac{\P^{\, 4}}{4!} + {\cal F} \frac{\P^{\, 2}}{\, 2!} + \frac{{\cal F}^2}{2} \Gamma^{11} \nn \\ \nn \\
&&\hspace*{-4.6cm}T_5 = \tau_1\frac{\P^{\, 5}}{5!}  + {\cal F} i \tau_2 \frac{\P^{\, 3}}{3!} + \frac{{\cal F}^2}{2} \tau_1 \P \nn 
\eea
\bea
&&\hspace*{-2.5cm}T_6 = \frac{\P^{\, 6}}{6!} + {\cal F} \Gamma^{11} \frac{\P^{\, 4}}{\, 4!} + \frac{{\cal F}^2}{2} \frac{\P^{\, 2}}{2} + \frac{{\cal F}^3}{3!} \Gamma^{11} \nn \\ \nn \\
&&\hspace*{-2.5cm}T_7 = i \tau_2 \frac{\P^{\, 7}}{7!} + {\cal F} \frac{\P^{\, 5}}{5!} \tau_1 + \frac{{\cal F}^2}{2} i \tau_2 \frac{\P^{\, 3}}{3!} + \frac{{\cal F}^3}{3!} \tau_1 \P \nn 
\eea
\bea
T_8 &=& \Gamma^{11} \frac{\P^{\, 8}}{8!} + {\cal F} \frac{\P^{\, 6}}{6!} + \frac{{\cal F}^2}{2} \Gamma^{11} \frac{\P^{\, 4}}{4!} + \frac{{\cal F}^3}{3!} \frac{\P^{\, 2}}{2} + \frac{{\cal F}^4}{4!} \Gamma^{11} \nn \\ \nn \\
T_9 &=& \tau_1 \frac{\P^{\, 9}}{9!} + {\cal F} i \tau_2 \frac{\P^{\, 7}}{7!} + \frac{{\cal F}^2}{2} \tau_1 \frac{\P^{\, 5}}{5!} + \frac{{\cal F}^3}{3!} i \tau_2 \frac{\P^{\, 3}}{3!} + \frac{{\cal F}^4}{4!} \tau_1 \P \,\, . \label{Tp}
\eea


\vfill\eject
  
\end{document}